\documentclass[12pt]{article}
\usepackage{amsmath,amssymb}
\begin{document}

\newtheorem{lem}{Lemma}[section]
\newtheorem{ex}{Example}[section]
\newtheorem{theorem}{Theorem}[section]
\newtheorem{prop}{Proposition}[section]
\newtheorem{rem}{Remark}[section]
\newtheorem{corollary}[theorem]{Corollary}
\newenvironment{proof}{\textbf{Proof:} }{\hfill $\blacksquare$}
\newtheorem{define}{Definition}[section]
\newtheorem{cor}{Corollary}[section]
\newtheorem{remark}{Remark}[section]
\allowdisplaybreaks

\makeatletter\@addtoreset{equation}{section}\makeatother
\def\theequation{\arabic{section}.\arabic{equation}}

\newcommand{\One}{1\!\!1}
\newcommand{\D}{{\cal D}}
\newcommand{\Sh}{{\cal S}}
\newcommand{\N}{{\Bbb N}}
\newcommand{\C}{{\Bbb C}}
\newcommand{\Z}{{\Bbb Z}}
\newcommand{\R}{{\Bbb R}}
\newcommand{\la}{\langle}
\newcommand{\ra}{\rangle}
\newcommand{\rom}[1]{{\rm #1}}
\newcommand{\FC}{{\cal F}{\cal C}_{\mathrm b}({\cal C}_0(\R^d),\Gamma)}
\newcommand{\eps}{\varepsilon}
\newcommand{\dd}{\overset{{.}{.}}}
\newcommand{\fii}{\varphi}

\newcommand{\supp}{\operatorname{supp}}

\def\stackunder#1#2{\mathrel{\mathop{#2}\limits_{#1}}}
\newcommand{\FCo}{{\cal F}{\cal C}_{\mathrm b}^\infty({\cal D},\dd\Gamma)}
\newcommand{\E}{{\cal E}}
\newcommand{\EG}{{\cal E}_{\mathrm G}}
\newcommand{\EK}{{\cal E}_{\mathrm K}}
\newcommand{\La}{\Lambda}
\newcommand{\ga}{\gamma} \newcommand{\Ga}{\square}

\renewcommand{\author}[1]{\medskip{\large #1}\par\medskip}
\begin{center}{\Large \bf
Spectral gap for Glauber type dynamics for a special class of
potentials.
}\end{center}

{\large Yuri Kondratiev}\\
 Fakult\"at f\"ur Mathematik, Universit\"at
Bielefeld, Postfach 10 01 31, D-33501 Bielefeld, Germany;
 BiBoS, Univ.\ Bielefeld,
Germany.\\ e-mail:
\texttt{kondrat@math.uni-bielefeld.de}\vspace{2mm}

{\large Tobias Kuna}\\ Department of Mathematics,
 University of Reading, Whiteknighs POBox~220, RG6~6AX Reading,
Berkshire, UK\\ e-mail:
\texttt{t.kuna@reading.ac.uk}\vspace{2mm}

{\large Nataliya Ohlerich}\\
 Fakult\"at f\"ur Mathematik, Universit\"at
Bielefeld, Postfach 10 01 31, D-33501 Bielefeld, Germany;
 BiBoS, Univ.\ Bielefeld,
Germany.\\ e-mail: \texttt{nturchyn@math.uni-bielefeld.de }

{\small

\begin{abstract}
We consider an equilibrium birth and death type process for a particle
system in infinite volume, the latter is described by
the space of all locally finite point
configurations on $\R^d$. These Glauber type dynamics are Markov
processes constructed for pre-given reversible measures.  A representation for the ``carr\'e du
champ'' and ``second carr\'e du champ'' for the
associate infinitesimal generators $L$ are calculated in infinite
volume and a corresponding coercivity identity is
derived. The latter is used to give explicit sufficient conditions for
the appearance and
bounds for the size of the spectral gap of $L$. These techniques are applied to Glauber
dynamics associated to Gibbs measure and conditions are derived
extending all previous known results. In the high temperature regime
now potentials also with a non-trivial negative part can be
treated. Furthermore, a special class of potentials is defined for
which the size of the spectral gap is as least as large as for the free
system and, surprisingly, is independent of
the activity. This type of potentials should not show any phase
transition for a
given temperature at any activity.
\end{abstract}

\noindent
 } \vspace{2mm}

\vspace{1.5mm}

 \vspace{1.5mm}

\section{Introduction}

The process studied in this paper is an analogue for continuous
systems of the well-known Glauber dynamics for lattice systems. The
main focus of the paper is on the spectral properties of the
associated infinitesimal generator $L$.
Such kind of dynamics were introduced for the first time by C.~Preston in
\cite{P,G1} for systems in finite volume, such that for each finite time
interval at most a finite number of particles appear in the system.
By construction, equilibrium states of classical statistical mechanics,
Gibbs measures, are formally reversible measures for such processes. Gibbs measures are perturbations of Poisson point
processes, though they are in general inequivalent to all Poisson point
processes, highly correlated and do not have necessarily nice decay of
correlation properties.
Gibbs measures are constructed using a pair potential $\phi$ and an
activity $z$.
In \cite{KL}, Yu.~Kondratiev and E.~Lytvynov
constructed the Glauber dynamics in infinite volume using Dirichlet-form
techniques. In any finite time interval, an infinite number of birth
and death events happen, therefore this process cannot be considered
as a birth and death process in the classical sense. In
infinite volume, the processes exist only in an $L^2$-sense with respect to a chosen
invariant measure $\mu$. For a more general construction in special
cases, see \cite{FKKZ10,LO08, KKM08a,KKZ06}.

The infinitesimal generator $L$ associated to these
dynamics have a spectral gap for
small positive potentials and small activity (high temperature regime).
In \cite{BCC}, L.~Bertini, N.~Cancrini and F.~Cesi derived a Poincare
inequality in finite volume and a bound on the spectral gap uniform in
the volume. They pointed out that typically a log-Sobolev type
inequality will not hold, cf. \cite{L99} for Poisson processes.
In \cite{KL} the technique of coercivity identity was used to improve the
result and to give a clear estimate for the spectral gap. In \cite{BCPP},
A.-S.~Boudou, P.~Caputo, P.~dai~Pra and G.~Posta derived a general
framework for this technique for general jump-type processes and
rederived the result for the Glauber dynamics in finite volume. In \cite{KMZ04},
Yu.~Kondratiev, R.~Minlos and E.~Zhizhina derive the one particle space invariant subspace and 
estimated the next gap in the spectrum.

In \cite{BE85}, D.~Bakry and M.~Emery calculated the ``second carr\'e
du champ'' generalizing the Bochner-Lich\'erowicz-Weitzenb\"ock formula and
in this way related the spectral gap of the
Laplacian on a manifold with the underlying curvature. Therefore, it
seems quite natural to apply these techniques also in the case of Glauber
dynamics in the continuum.

In Section~\ref{rfyje}, we consider, slightly
more general, all measure which have an integration by
parts formula with respect to the considered difference operator, in
other words measures which have a Papangelou kernel.
We calculate the ``second carr\'e
du champ'' in infinite
volume under very mild assumptions on the invariant measure $\mu$ and
the associated Papangelou kernel exploiting fundamentally the
pointwise nature of the ``second carr\'e
du champ''. We recover in an equivalent form the coercivity identity
given in \cite{KL} and exactly the one given in \cite{BCPP}, however in
infinite volume. This technique has the advantage to provide a
motivation which particular form of the coercivity identity to use, although a
geometrical justification could not be given. However, the results
presented in this paper may motivate
an adequate
geometrical structure on configuration spaces. Sufficient criteria
for the presence of a spectral gap are derived from the coercivity identity.

In Section~\ref{secgibbs}, we study the case of operators $L$ associated
to Gibbs measures in more details.
Sufficient conditions for the presence of a spectral gap are derived
 and bounds on the size of the gap in terms of
the potential and the activity are given. We introduce a class of
non-trivial potentials for which the spectral gap has at least the size as in the free case and, even
more surprisingly, the derived bound on the size of the spectral gap is
independent of the
activity.  The definition of this class is based upon Fourier transform and hence
the continuous space structure of the system is essential.
Even more surprisingly, there are potentials with
non-trivial negative part in this class. Furthermore, do we show that
an increase in the
temperature will not alter these estimates as well.
For positive potentials from this class, this result improves
essentially the bound given in \cite{KL}.

Finally, we derive
a bound for potentials which are the sum of a potential from the
aforementioned special class and a usual regular and stable potential
in a generalized high temperature regime. The size of the spectral gap is
estimated in terms of the density and not of the activity, which is
more satisfying from the viewpoint of physics.
This result gives, in particular, an improved estimate on the size of
the spectral gap even if one just considers a generic stable and regular
potentials alone. Till now only non-negative potentials could be
treated and even for general positive potentials the previous
results are improved.

Precisely speaking we do not derive a spectral gap but a coercivity
inequality on cylinder functions. If $L$ is essentially self-adjoint
on this domain, as proven for positive potentials in \cite{KL}, then
the coercivity identity is equivalent to spectral gap. Essential
self-adjointness for non-positive potentials will be subject of future
investigations.

Assuming essential self-adjointness, we found a class of potentials
with a very interesting thermodynamical property. These potentials
have a non-trivial attractive part, nevertheless there will be no phase
transition of any kind for all values of the activity $z$.

\section{States and dynamics}\label{fsc}

\subsection{Configuration space}
The configuration space $\Gamma:=\Gamma_{\R^d}$ over $\R^d$ is
defined as the set of all Radon measures with values in $\mathbb{N}
\cup \{0,\infty\}$, i.e.
for any $\gamma \in \Gamma$ there exists a sequence $(x_i)_{i \in I}$ of vectors
from $\R^d$ and an index set $I \subset \mathbb{N}$  such that $\gamma =
\sum_{i\in I} \eps_{x_i}$.
Conversely, any sequence
without accumulation points can be associated to a configuration by the above formula.
Modulo renumeration there is only one sequence representing $\gamma$.
The space $\Gamma$ is Polish in the relative topology as a subset of
the space off all Radon measures
${\cal M}(\R^d)$ endowed with the vague topology, i.e. the topology generated by the mappings
\[
\ga \mapsto \langle f, \ga \rangle := \int_{\R^d} f(x) \ga(dx) \qquad \mathcal{C}_0(\R^d),
\]
where $\mathcal{C}_0(\R^d)$ denotes the set of all continuous functions on $\R^d$ with compact support.
 The corresponding Borel $\sigma$-algebra on $\Gamma$ is denoted by ${\cal B}(\Gamma)$.
A probability measure on $(\Gamma,{\cal B}(\Gamma))$ is called a point
process (random field).
A measurable function $r:\R^d\times \Gamma\longrightarrow [0,\infty]$ is
the Papangelou intensity of a point process $\mu$ if
\begin{equation}
\int_\Gamma \mu(d\gamma)\int_{\R^d} \gamma(dx)
F(x,\gamma) =\int_\Gamma \mu(d\gamma)\int_{\R^d}
dx\cdot r(x,\gamma) F(x,\gamma +\delta_x)\label{mecke}
\end{equation}
for any measurable function $F:\R^d\times \Gamma \to[0,+\infty[.$
Let us fix a point process $\mu$ which has Papangelou intensity $r$
and for which the first correlation function exists. The first $n$
correlation functions exists exactly iff $\mu$ has  all local moments
up to degree $n$, that is, for all bounded measurable subsets $\La \subset \R^d$ the following
integral $\int_{\Gamma} \ga(\La)^n \mu(d\ga)$ is finite.

\subsection{Glauber dynamics}

In this subsection we introduce the Glauber dynamics, a birth and death
type dynamics in the continuum via Dirichlet form techniques, for
details cf. \cite{KL}.
For this purpose we first introduce the set $\FC$
of all functions of the form $$\Gamma\ni \gamma\mapsto F(\gamma)=g_F(\la\varphi_1,\gamma\ra,\dots , \la\varphi_N,\gamma\ra), $$ where $N\in\N$, $\varphi_1,\dots,\varphi_N\in \mathcal{C}_0(\R^d)$ and
$g_F\in \mathcal{C}_{\mathrm b}(\R^N)$. Here $\mathcal{C}_{\mathrm b}(\R^N)$ denotes the set of all continuous
bounded functions on $\R^N$. The dynamics is constructed using two
types of difference operators which are in some sense adjoint to each other:
for $F:\Gamma\to\R$, $\gamma\in\Gamma$, and $x,y\in \R^d$
\begin{equation}\label{DxDy}
(D_x^-F)(\gamma):=F(\gamma - \delta_x)-F(\gamma),\quad
(D_x^+F)(\gamma):=F(\gamma) - F(\gamma + \delta_x).\end{equation}
As we want to consider the dynamics only in an $L^2$-framework, we
use the following  bilinear form, cf.  \cite{KL}
\begin{align} \E(F,G):=& \int_\Gamma\mu(d\gamma)\int_{\R^d} \gamma(dx) (D_x^-F)(\gamma) (D_x^-G)(\gamma),
\qquad F,G\in\FC,\label{1}
\end{align}
The following properties of the $\E$, which are useful for our considerations, where proved in \cite{KL}.

Using the associated integration by parts formula for a measure $\mu$ with a
Papangelou intensity $r$ and first local moments in \cite{KL} it was proven that the bilinear
form $(\E,\FC)$ is closable on $L^2(\Gamma,\mu)$ and its closure is a Dirichlet form also
denoted by $(\E,D(\E))$. The generator $(L,D(L))$ associated to$(\E,D(\E))$, i.e. $\E(F,G)=(-LF,G)_{L^2(\Gamma,\mu)}$ is for
functions $F\in\FC \subset D(L)$ given by
\begin{equation}\label{genG}
(LF)(\gamma)=\int_{\R^d}\gamma(dx)\,(D^-_xF)(\gamma)-\int_{\R^d}
 r(x,\gamma) (D^+_xF)(\gamma)dx\qquad \text{\rom{$\mu$-a.e.}}.
\end{equation}
Following the usual techniques for Dirichlet forms, in \cite{KL}, for the case,
that $\mu$ is a Gibbs measure,
the associated conservative Hunt process was constructed, that is,
$${\bf M}=({\pmb{ \Omega}},{\bf F},({\bf F}_t)_{t\ge0},({\pmb
\Theta}_t)_{t\ge0}, ({\bf X}(t))_{t\ge 0},({\bf P
}_\gamma)_{\gamma\in\Gamma})$$ on $\Gamma$ \rom(see e\rom.g\rom.\
\rom{\cite[p.~92]{MR})} which is properly associated with $({\cal
E},D({\cal E}))$\rom, i\rom.e\rom{.,} for all \rom($\mu$-versions
of\/\rom) $F\in L^2(\Gamma,\mu)$ and all $t>0$ the function
\[ \Gamma\ni\gamma\mapsto
p_tF(\gamma){:=}\int_{\pmb\Omega} F({\bf X}(t))\, d{\bf
P}_\gamma
\]
 is an ${\cal E}$-quasi-continuous version
of $\exp(tL)F$\rom. $\Omega$ is the set of all \textit{cadlag}
functions $[0,\infty[ \rightarrow \Gamma$. The processes $\bf M$ is up to $\mu$-equivalence unique
\rom(cf\rom.\ \rom{\cite[Chap.~IV, Sect.~6]{MR}).} In
particular\rom, ${\bf M}$ is $\mu$-symmetric \rom(i\rom.e\rom{.,}
$\int G\, p_tF\, d\mu=\int F \, p_t G\, d\mu$ for all
$F,G:\Gamma\to{\Bbb R}_+$\rom, ${\cal B}(\Gamma)$-measurable\rom)\rom, and
 thus has $\mu$ as an invariant measure\rom.

\section{Coercivity identity for Glauber dynamics}\label{rfyje}

\subsection{Carr\'e du champ}\label{subseccarre}

In this subsection we compute two quadratic forms associated to $L$,
the generator of Glauber dynamics given by \eqref{genG}, the so-called
``carr\'e du champ'',
the ``second carr\'e du champ'' and hence an analogue of the
Bochner-Lichn\'erowicz-Weitzenb\"ock formula in this context, cf. e.g. \cite{Bak94}. As this is essentially an algebraic calculation most details
are omitted and we give just the main steps of these computation, which should
allow the interested reader to easily reconstruct the missing details.

In this subsection we essentially need only the following assumption
on $r: \R^d \times \Gamma \rightarrow [0,\infty]$: There exists a
subset $\Gamma_{\mathrm{temp}} \subset \Gamma$ such that
\begin{enumerate}
\item $r(x,\gamma) < \infty$ for
all $(x,\gamma) \in \R^d \times \Gamma_{\mathrm{temp}}$
\item for all $\gamma \in \Gamma_{\mathrm{temp}}$, the function
$x \mapsto r(x,\gamma)$ is locally integrable
\item for all
$\gamma \in \Gamma_{\mathrm{temp}}$ and all $x\in \gamma$ and $ y \in
\R^d$ also $\gamma -\delta_x$ and $\gamma +\delta_y$ are in
$\Gamma_{\mathrm{temp}}$.
\end{enumerate}
For $F,G \in \FC$ we define the ``carr\'e du champ'' corresponding to $L$ as
\begin{equation}\label{carre}
\Ga(F,G):=\frac12(L(FG)-FLG-GLF).
\end{equation}
Let us split the generator $L$ into its death and birth part
\begin{equation}
L^-F(\gamma) :=  \sum_{x \in \gamma} D^-_xF(\gamma), \qquad
L^+F(\gamma) := \int_{\R^d} r(x,\gamma) D_x^+F(\gamma)dx,
\end{equation}
such that $L =L^- - L^+$. Due to linearity one obtains that $\Ga(F,G)=\Ga^-(F,G)+\Ga^+(F,G)$, where
$\Ga^-$ and $-\Ga^+$ are the ``carr\'e du champ'' corresponding to the death and birth parts
$$\Ga^-(F,G):=\frac12\int_{\R^d} \ga(dx)D_x^- F(\ga)D_x^- G(\ga),\,\,\,\,\,
\Ga^+(F,G):=\frac12\int_{\R^d} r(x,\ga)D_x^+ F(\ga)D_x^+ G(\ga)dx.$$

Iterating in some sense the definition of ``carr\'e du champ'' one may introduce the so-called
$\Ga_2$, cf. \cite{Bak94},  as follows
\begin{equation}\label{carre2}
2 \Ga_2(F,F):= L\Ga(F,F)-2\Ga(F,LF).
\end{equation}
The splitting in birth and death part allows us to split $\Ga_2$ correspondingly in the following way:
\begin{eqnarray}
2 \Ga_2(F,F)&=&  \left( L^-\Ga^{-}(F,F) - 2 \Ga^-(F,L^-F) \right) \label{eq2}\\
&&- \left( L^+\Ga^{+}(F,F) - 2 \Ga^+(F,L^+F) \right) \notag\\
&&+  \left( L^-\Ga^{+}(F,F)  - 2 \Ga^+(F,L^-F)  \right) \notag\\
&&- \left( L^+\Ga^{-}(F,F) - 2 \Ga^-(F,L^+F) \right)
\nonumber
\end{eqnarray}
All brackets  will be calculated separately using the following product rules type formulas
\begin{lem}\label{rules}
If $H: \R^d \times \Gamma_{\mathrm{temp}} \rightarrow \R$ is locally
bounded and for fixed $\gamma \in \Gamma_{\mathrm{temp}}$ the function
$x \mapsto H_x(\gamma)$ has compact support, then
\begin{eqnarray}
D_x^+ \sum_{y \in \gamma} H_y(\gamma) &=& \sum_{y \in \gamma}
D_x^+H_y(\gamma)-H_x(\gamma + \delta_x) \\
D_x^- \sum_{y \in \gamma} H_y(\gamma) &=& \sum_{y \in \gamma -\delta_x}
D_x^-H_y(\gamma)-H_x(\gamma )  \\
D_x^+\left(\int_{\R^d} r(y,\ga)H_y(\ga)dy\right)\!\!\!\!&=&\!\!\!\! \int_{\R^d}\!\!\!\! r(y,\ga)D_x^+H_y(\ga)dy+
\int_{\R^d}\!\!\!\! D_x^+r(y,\ga)H_y(\ga +\delta_x)dy, \label{D+int} \\
D_x^-\left(\int_{\R^d} r(y,\ga)H_y(\ga)dy\right)\!\!\!\! &=&\!\!\!\! \int_{\R^d}\!\!\!\! r(y,\ga)D_x^-H_y(\ga)dy+
\int_{\R^d}\!\!\!\! D_x^-r(y,\ga)H_y(\ga - \delta_x)dy. \label{D-int}
\end{eqnarray}
\end{lem}
Computing the first summand of (\ref{eq2}) we obtain
\[
L^-\Ga^-(F,F)(\gamma) -2 \Ga^-(L^-F,F)(\gamma)
=  \frac{1}{2}\sum_{x \in \gamma}\sum_{y \in \gamma -\delta_x}
\left( D_x^-D_y^-F\right)^2(\gamma)
+   \Ga^-(F,F)(\gamma),
\]
whereas for the second summand we may derive the following expression
\begin{eqnarray*}
&&L^+\Ga^+(F)(\gamma)-2\Ga^+(F,L^+F)(\gamma) =
 - \frac12\int_{\R^d}\int_{\R^d}  r(x,\ga) r(y,\ga) (D_x^+D_y^+ F)^2(\ga) dx dy\\
&& + \frac12\int_{\R^d} \int_{\R^d} r(x,\ga) D_x^+ r(y,\cdot)(\ga)(D_y^+ F)^2(\ga + \delta_x) dxdy\\
&& - \int_{\R^d} \int_{\R^d} r(x,\ga) D_x^+ F(\ga) D_x^+ r(y,\cdot)(\ga)D_y^+F(\ga+\delta_x)dxdy
\end{eqnarray*}
Finally, calculating the mixed terms in (\ref{eq2}), we obtain
\begin{eqnarray*}
 (L^-\Ga^+(F)-2\Ga^+(F,L^-F)  &=& \sum_{x\in\ga}\int_{\R^d} r(y,\ga) (D_x^-D_y^+ F)^2 (\ga) dy
\\
&& + \frac12\sum_{x\in\ga}\int_{\R^d} D_x^- r(y,\cdot)(\ga)(D_y^+ F)^2(\ga - \delta_x)dy\\
-L^+\Ga^-(F)+2\Ga^-(F,L^+F))(\gamma)& =& \sum_{x\in\ga}\int_{\R^d} r(y,\ga) (D_x^-D_y^+ F)^2 (\ga) dy
\\ && + \frac12\int_{\R^d} r(y,\ga)(D_y^+ F)^2(\ga)dy\\
&& +\sum_{y\in\ga}D_y^- F(\ga)\int_{\R^d}
D_y^-r(x,\cdot)(\ga)D_x^+F(\ga -\delta_y)dx
\end{eqnarray*}
Summarizing, adding all four parts we gain the following expression for $\Ga_2$
\begin{eqnarray}
\lefteqn{\Ga_2(F,F)(\gamma)}\\ &=& \frac{1}{2}\Ga(F,F)(\gamma) + \Ga(F,F)^+(\gamma)\label{thseconcarre}\\\nonumber
&+ &\frac{1}{4} \sum_{x \in \gamma}\sum_{y \in \gamma - \delta_x}
\left( D_x^-D_y^-F\right)^2(\gamma) + \frac{1}{2} \sum_{y \in \gamma} \int_{\R^d} r(x,\gamma )
\left(D_x^+ D_y^-F\right)^2\!(\gamma ) dx\\\nonumber
&+& \frac{1}{4}\int_{\R^d} \sum_{x \in \gamma} D_x^-r(y,\cdot)(\gamma) \left[(D_y^+F)^2(\gamma -\delta_x)
+2 D_y^+F(\gamma -\delta_x) D_x^-F(\gamma) \right]dy \nonumber \\
&+&\frac{1}{4}\int_{\R^d}  \int_{\R^d} r(x,\gamma)r(y,\gamma) (D_x^+D_y^+F)^2(\gamma)dxdy\nonumber\\
&+&\frac{1}{4}\int_{\R^d} r(x,\gamma) \int_{\R^d} D_x^+r(y,\cdot)(\gamma) \left[-(D_y^+F)^2(\gamma+\delta_x)
+2 D_y^+F(\gamma+\delta_x)D_x^+F(\gamma) \right]dydx\nonumber
\end{eqnarray}
This representation is still not in a convenient form. For Gaussian
type measures there is a Bochner-Lichn\'erowicz-Weitzenb\"ock kind
formula and an associated Bakry-Emery criterium for
$\Ga_2$ in terms of geometrical quantities like
the underlying curvature and the Hessian. Unfortunately, in our
case we lack this understanding of the associated geometrical
structure.
However, we observe that we have three terms of fourth order
in the differential operator. One may expect that in a natural
representation they all would have
all the same integral w.r.t. the reversible measure $\mu$, which, as
we will see, is not the case
for the second but last summand, cf. (\ref{eq3}). Therefore, we rearrange the last and second but last
summand in (\ref{thseconcarre}) and obtain the following
Bochner-Lichn\'erowicz-Weitzenb\"ock formula
\begin{theorem}
\label{thmnsecondcarre}
For all $F,G \in \FC$ it holds that
\begin{eqnarray}
\Ga_2(F,F)(\gamma)&=& \frac{1}{2}\Ga(F,F)(\gamma) + \Ga^+(F,F)(\gamma)\nonumber\\\nonumber
&+ &\frac{1}{4} \sum_{x \in \gamma}\sum_{y \in \gamma -\delta_x}
\left( D_x^-D_y^-F\right)^2(\gamma) + \frac{1}{2} \sum_{y \in \gamma} \int_{\R^d} r(x,\gamma )
\left(D_x^+ D_y^-F\right)^2\!(\gamma ) dx\\\nonumber
&+& \frac{1}{4}\int_{\R^d} \sum_{x \in \gamma} D_x^-r(y,\cdot)(\gamma) \left[(D_y^+F)^2(\gamma -\delta_x)
+2 D_y^+F(\gamma -\delta_x) D_x^-F(\gamma) \right]dy \nonumber \\
&+&\frac{1}{4}\int_{\R^d} r(x,\gamma) \int_{\R^d} r(y,\gamma +\delta_x ) (D_x^+D_y^+F)^2(\gamma)dydx\nonumber\\
&+&\frac{1}{4}\int_{\R^d} r(x,\gamma) \int_{\R^d} D_x^+r(y,\cdot)(\gamma) \left[-(D_y^+F)^2(\gamma)+2
D_y^+F(\gamma)D_x^+F(\gamma) \right] dydx\nonumber
\end{eqnarray}
\end{theorem}

\begin{proof} Using just the definition of $D_x^+$ the last two
  summand in (\ref{thseconcarre}) can be rewritten as follows
\begin{eqnarray}
\lefteqn{ \frac{1}{4}\int_{\R^d} r(x,\gamma) \int_{\R^d} r(y,\gamma +\delta_x ) (D_x^+D_y^+F)^2(\gamma)dydx} \label{eq1} \\
&&\hspace{-0.5cm}+\frac{1}{4}\int_{\R^d}\!\!\!\! r(x,\gamma) \int_{\R^d}\!\!\!\! D_x^+r(y,\cdot)(\gamma) \left[(D_x^+D_y^+F)^2(\gamma)-
(D_y^+F)^2(\gamma +\delta_x)
+2 D_y^+F(\gamma +\delta_x)D_x^+F(\gamma) \right]dydx \nonumber
\end{eqnarray}
It remains to simplify the last bracket. Expanding the first summand of the bracket and
using
$$-D_x^+(D_y^+F)^2(\gamma)=(D_y^+F)^2(\gamma +\delta_x)-(D_y^+F)^2(\gamma)$$
$$2D_x^+(D_y^+F)(\gamma)D_y^+F(\gamma)=2(D_y^+F(\gamma)-D_y^+F(\gamma + \delta_x))D_y^+F(\gamma)$$
we get
\begin{eqnarray*}
\lefteqn{(D_x^+D_y^+F)^2(\gamma)-(D_y^+F)^2(\gamma +\delta_x)+2 D_y^+F(\gamma +\delta_x)D_x^+F(\gamma)}\\
&=& - D_x^+(D_y^+F)^2(\gamma) + 2 D_x^+D_y^+F(\gamma) D_y^+F(\gamma)-
(D_y^+F)^2(\gamma + \delta_x)
+2 D_y^+F(\gamma +\delta_x)D_x^+F(\gamma)\\
&=& - (D_y^+F)^2(\gamma) + 2 D_x^+F(\gamma) D_y^+F(\gamma)
-2 D_x^+F(\gamma + \delta_y) D_y^+F(\gamma)+2 D_y^+F(\gamma +\delta_x)D_x^+F(\gamma)
\end{eqnarray*}
According to Lemma~\ref{lemsym} below the integral expression w.r.t. which one has
to integrate the afore calculated summand is symmetric under the interchange of $x$ and $y$.
Hence the last two terms in the previous calculation cancel each other and the second
summand in (\ref{eq1}) can be simplified to
\begin{eqnarray*}
\int_{\R^d} r(x,\gamma) \int_{\R^d} D_x^+r(y,\cdot)(\gamma)[- (D_y^+F)^2(\gamma) +
2 D_x^+F(\gamma) D_y^+F(\gamma)]dydx,
\end{eqnarray*}
which yields the result.
\end{proof}

\begin{lem}
\label{lemsym}
For $\mu \otimes dx$-a.a. $(\gamma,x)$ holds that
\[
r(x,\gamma) D_x^+r(y,\cdot)(\gamma)dxdy = r(y,\gamma) D_y^+r(x,\cdot)(\gamma)dydx
\]
\end{lem}

\begin{proof}
As the above equality has to be interpreted a.s. it is sufficient to show that the
following expression is invariant under the interchange of $x$ and $y$
for any cylinder function $H$. This is
obvious after the following rewriting
\begin{eqnarray*}
\lefteqn{\int_\Gamma \int_{\R^d} r(x,\gamma) \int_{\R^d} D_x^+r(y,\cdot)(\gamma) H(\gamma + \delta_x +\delta_y,x,y) dydx
\mu(d\gamma)}\\
&=& \int_\Gamma \int_{\R^d} r(x,\gamma) \int_{\R^d} r(y,\gamma) H(\gamma + \delta_x +\delta_y,x,y) dydx\mu(d\gamma)\\
&&- \int_\Gamma \sum_{ \stackrel{\scriptstyle x,y \in \gamma}{x \neq y} }
H(\gamma ,x,y) \mu(d\gamma)
\end{eqnarray*}
\end{proof}

\subsection{Coercivity identity}

In order to study spectral properties of $L$ we consider integrals of
$\Ga$ and
$\Ga_2$ with respect to an associated probability $\mu$, that is a
probability measure  with a Papangelou
intensities $r$, cf. \eqref{mecke}.  The representation given in Theorem \ref{thmnsecondcarre}
yields a particular representation useful for this purpose.

In this subsection we need to  assume that $\mu$ has not only local
moments up to first but up to second order.
In particular, then  for all compact $\Lambda \subset
\R^d$ holds that $\gamma \mapsto \int_\Lambda
\int_\Lambda
r(y,\gamma) r(y,\gamma +\delta_x)$ is integrable w.r.t $\mu$.
In order that $\FC \subset D(L^2)$, we have additionally to assume that
$\gamma \mapsto \int_\Lambda r(x,\gamma) dx $ is in $L^2(\Gamma,
\mu)$. Then $r$ has a version which fulfills all assumptions used in
Subsection~\ref{thmnsecondcarre} hold for a set
$\Gamma_{\mathrm{temp}}$ of full measure.

Recall that $L$ is symmetric with respect to
$\mu$ and $L$ applied to constant functions is zero.
Using that we get the following relations for $\Ga$ and $\Ga_2$:
for all $F \in \FC$ holds
\begin{eqnarray*}
\E(F,F)=\int_{\Gamma} F(\gamma) LF(\gamma) \mu(d\gamma) &=&\int_{\Gamma} \Ga(F,F)(\gamma) \mu(d\gamma).
\\
\int_{\Gamma} (LF)^2(\gamma) \mu(d\gamma) &= &\int_{\Gamma} \Ga_2(F,F)(\gamma)
\mu(d\gamma)
\end{eqnarray*}

The following identities are derived using repeatedly the identity
$D_x^+F(\gamma -\delta_x)=D_x^-F(\gamma)$.
and the definition of the Papangelou intensities, cf. \eqref{mecke}.
For all $F \in \FC$ holds
$$\frac12\int_{\Gamma}\Ga(F,F)(\ga)\mu(d\ga) =\int_{\Gamma}\Ga^\pm(F,F)(\ga)\mu(d\ga)=
\int_{\Gamma} \int_{\R^d} r(x,\ga)(D_x^+F)^2(\gamma) dx \mu(d\gamma)
$$
and, in particular, one gets the representation (\ref{1}) for
the Dirichlet form $\E$.

Furthermore, in the representation given in Theorem~\ref{thmnsecondcarre} the expectations
of all fourth order terms coincides, that is,
for all $F \in \FC$ holds
\begin{eqnarray*}
\lefteqn{\int_\Gamma \sum_{y \in \gamma} \int_{\R^d} r(x,\gamma )
\left(D_x^+ D_y^-F\right)^2\!(\gamma ) dx\mu(d\gamma)}\\
&=& \int_\Gamma \sum_{x \in \gamma}\sum_{y \in \gamma -\delta_x }
\left(D_x^+ D_y^-F\right)^2\!(\gamma -\delta_x ) \mu(d\gamma)\\
&=& \int_\Gamma \sum_{x \in \gamma}\sum_{y \in \gamma -\delta_x }
\left(D_x^- D_y^-F\right)^2\!(\gamma) \mu(d\gamma).
\end{eqnarray*}
and indeed in Subsection~\ref{subseccarre} the last fourth order term in (\ref{thseconcarre})
was rearranged in such a form in Theorem~\ref{thmnsecondcarre} that now holds
\begin{equation}\label{eq3}
\int_\Gamma \int_{\R^d} r(x,\gamma) \int_{\R^d} r(y,\gamma +\delta_x ) (D_x^+D_y^+F)^2(\gamma) dydx\mu(d\gamma)
= \int_\Gamma \sum_{y \in \gamma}\sum_{x \in \gamma -\delta_y}  (D_x^-D_y^-F)^2(\gamma)\mu(d\gamma)
\end{equation}
For the remaining second order terms in Theorem~\ref{thmnsecondcarre} one can find some
cancelations. For all $F \in \FC$ holds
\begin{eqnarray}
\lefteqn{\int_\Gamma
\int_{\R^d} \sum_{x \in \gamma} D_x^-r(y,\cdot)(\gamma) \left[(D_y^+F)^2(\gamma -\delta_x)
+2 D_y^+F(\gamma -\delta_x) D_x^-F(\gamma) \right]dy\mu(d\gamma)} \nonumber\\
&=&\int_\Gamma \int_{\R^d} r(x,\gamma) \int_{\R^d} D_x^-r(y,\cdot)(\gamma +\delta_x) \left[(D_y^+F)^2(\gamma)
+2 D_y^+F(\gamma ) D_x^-F(\gamma +\delta_x) \right]dydx \mu(d\gamma)\nonumber \\
&=&\int_\Gamma \int_{\R^d} r(x,\gamma) \int_{\R^d} D_x^+r(y,\cdot)(\gamma ) \left[(D_y^+F)^2(\gamma)
+2 D_y^+F(\gamma ) D_x^+F(\gamma ) \right]dydx \mu(d\gamma).\nonumber
\end{eqnarray}
Note that the first summand in the last term has the opposite sign as
the first summand in the
last term of the representation given in Theorem~\ref{thmnsecondcarre}.
Summarizing one obtains the coercivity identity
\begin{theorem}\label{Ga2}
For all $F \in \FC$ holds that
\begin{eqnarray*}
\int_{\Gamma} (LF)^2(\gamma) \mu(d\gamma) &=&\int_{\Gamma} \Ga_2(F,F)(\gamma) \mu(d\gamma)\\
&=& \int_{\Gamma} \Ga(F,F)(\gamma) \mu(d\gamma)
+  \int_{\Gamma} \sum_{x \in \gamma}\sum_{y \in \gamma -\delta_x} \left( D_x^-D_y^-F\right)^2(\gamma)
\mu(d\gamma)\notag \\
&+& \int_\Gamma \int_{\R^d} r(x,\gamma) \int_{\R^d} D_x^+r(y,\cdot)(\gamma)
D_y^+F(\gamma)D_x^+F(\gamma)dydx \mu(d\gamma). \notag
\end{eqnarray*}
\end{theorem}

\subsection{Sufficient condition for spectral gap}

Instead of proving spectral gap directly using the Poincare
inequality, we consider the following approach, see \cite{K85} and
\cite[Chapter.~6, Section~4]{BK}.

Let $L$ be a nonnegative self-adjoint operator which maps the constant
functions to  zero. Let $D(L)$ be a core of $L$ and $c >0$.
Then $L$ has a spectral gap of at least $c$
if and only if the following so-called coercivity inequality holds
\begin{equation}\label{poi}
\int_{\Gamma}(LF)^2(\ga)\mu(d\ga)\geq c \E(F,F), \quad \forall F \in D(L).
\end{equation}
The latter inequality can be expressed in terms
of the ``carr\'e du champ'' $\Ga$ and $\Ga_2$
\begin{equation}\label{poiGa}
\int_{\Gamma}\Ga_2(F,F)(\ga)\mu(d\ga)\geq c\int_{\Gamma} \Ga(F,F)(\ga)\mu(d\ga).
\end{equation}
For diffusions D.~Bakry and M.~Emery could derive directly an inequality for $\Ga$ and $\Ga_2$, cf. \cite{BE85},
which we are not able to do.

Inserting in \eqref{poiGa} the representations of the previous sections one easily derives the following
inequality
\begin{eqnarray}
\lefteqn{(1-c)\int_{\Gamma}\int_{\R^d} r(x,\ga)(D_x^+F)^2(\gamma) dx\mu(d\gamma)}\label{cond} \\
&+& \int_{\Gamma} \int_{\R^d} \int_{\R^d}
r(x,\gamma)D_x^+r(y,\cdot)(\gamma)D_y^+F(\gamma)D_x^+F(\gamma)dydx
\mu(d\gamma) \geq 0 \nonumber
\end{eqnarray}
using that
\[
\sum_{x \in \gamma}\sum_{y \in \gamma -\delta_x} \left(
  D_x^-D_y^-F\right)^2(\gamma)
\geq 0
\]
Considering the integrand \eqref{cond} for fixed $\gamma$ and denoting by
$$K_{\ga}(x,y)=r(x,\ga)(r(y,\ga)-r(y,\ga +\delta_x)),\,\,\, \psi_{\ga}(x)=D_x^+F(\ga).$$
we can give a sufficient condition for the inequality \eqref{cond}
to hold for all $F \in \FC$, namely
for all $\psi \in \mathcal{C}_0(\R^d)$ holds
\begin{equation}\label{posdefv1}
\int_{\R^d} \int_{\R^d} (K_{\ga}(x,y)+(1-c) \sqrt{r(x,\ga)}\sqrt{r(y,\ga)}\delta(x-y))\psi(y)\psi(x) dxdy \geq 0.
\end{equation}
This can be formulate more elegantly
using the following definition
\begin{define}
A Radon measure $K$ on $\R^d\times\R^d$ is called a positive definite kernel
if for all $\psi\in \mathcal{C}^\infty_0(\R^d)$ holds
\begin{equation}
\int_{\R^d}\int_{\R^d}  \psi(x)\psi(y)K(dx,dy) \geq 0
\end{equation}
\end{define}

\begin{theorem}\label{thsuff}
If there is a $c >0 $ such that for $\mu$-a.a. $\ga$ the kernel
\begin{equation}\label{posdef}
r(x,\ga)(r(y,\ga)-r(y,\ga +\delta_x))+(1-c) \sqrt{r(x,\ga)}\sqrt{r(y,\ga)}\delta(x-y)
\end{equation}
is positive definite then the coercivity inequality \eqref{poi} for $L$
with constant $c$ holds for all $F \in \FC$.
\end{theorem}

\section{Coercivity identity for Gibbs measures}\label{secgibbs}

In this section we demonstrate that the sufficient condition for the
coercivity inequality developed in Theorem~\ref{thsuff} gives surprising results
for the Glauber dynamics associated to Gibbs measures.

\subsection{Gibbs measures}

Gibbs measures are just the measures with Papangelou
intensities of the form
$r(x,\ga)=z \exp[-E(x,\ga)]$, where $z > 0$ and
\[
E(x,\ga):=\left\{
\begin{array}{ll}
\sum\limits_{y\in\ga}\phi(x-y),& \text{ if } \sum\limits_{y\in\ga}|\phi(x-y)|<\infty,\\
+\infty, & \text{ otherwise, }
\end{array}
\right.
\]
for a measurable symmetric function $\phi : \R^d \to (-\infty,\infty]$.
One calls such a measure a Gibbs measure to the activity $z$ and pair
potential $\phi$. Sometimes it is useful to introduce an extra
parameter, the inverse temperature $\beta$, and consider Gibbs
measures for $\beta \phi$.

To guarantee existence of a measure with such Papangelou intensities, we need to require
further conditions on the pair potential $\phi$.
For every $r\in\Z^d$, define a cube $\Delta_r=\left\{x\in\R^d: r_i-\frac12\leq x_i<r_i+\frac12\right\}.$
These cubes form a partition of $\R^d.$ Denote by
$N_r(\ga)=\ga(\Delta_r).$ One says that $\phi$ is superstable (SS) if there exist
$A>0,$ $B\ge0$ such that, for all $\gamma\in\Gamma$ such that $\gamma(\R^d)<\infty$ holds
$$\sum_{\{x,y\}\subset\gamma}\phi(x-y)\ge
\sum_{r\in\Z^d}AN_r^2(\ga)-BN_r(\ga).$$ $\phi$ is called stable (S) if
the above condition
holds just for $A=0$. One says that $\phi$ is regular (R) if $\phi$ is
bounded below and
there exists an $R>0$ and a positive decreasing function $\varphi$
on $[0,+\infty)$ such that $|\phi(x)|\le\varphi(|x|)$ for all  $x\in\R^d$ with $|x|\ge R$ and
\begin{equation}\label{eq4}\int_R^{\infty}t^{d-1}\varphi(t)dt<\infty.\end{equation}
For the notion of tempered Gibbs measure and the
following theorem, see \cite{Ru70}.

\begin{theorem}
\label{yte6te645}

Let $\phi$ be (SS) and (R), then the set ${\cal
G}_{{\mathrm temp}}(z,E)$ of all tempered  Gibbs measures is
non-empty and for each measure from ${\cal G}_{{\mathrm temp}}(z,E)$
all correlation functions exist and satisfy the so-called Ruelle bound\rom.
\end{theorem}

A Gibbs measure that fulfills the Ruelle bound has all (local) moments
and one can see quite easily that also $\gamma \mapsto \int_{\R^d} r(x,\gamma) dx $ is in $L^2(\Gamma,
\mu)$, cf. e.g. \cite{KL}. Hence all assumptions of Subsection~\ref{subseccarre}
are fulfilled. Hence, in the sequel, we will restrict ourself to Gibbs measures which
fulfill a Ruelle bound.

\subsection{Coercivity inequality}

For Gibbs measures condition~\eqref{posdef} takes the following form

\begin{theorem}
Let $\mu$ be a Gibbs measure for a pair potential $\phi$ and activity
$z$ which fulfills a Ruelle bound. If for
a.a. $\ga$ the kernel
\begin{equation}\label{Gibbsposdef}
e^{-E(x,\ga)}e^{-E(y,\ga)}z(1-e^{-\phi(x-y)})+(1-c) e^{-\frac12 E(x,\ga)}e^{-\frac12 E(y,\ga)}\delta(x-y)
\end{equation}
is positive definite then the coercivity inequality \eqref{poi} for $L$
with constant $c$ holds for all $F \in \FC$.
\end{theorem}

The following easy reformulation will become very fruitful later on.
Using in  \eqref{posdefv1} the function $e^{-\frac12E(x,\ga)} \psi(x)$ instead of $\psi$
gives
\begin{corollary}
Let $\mu$ be a Gibbs measure for a pair potential $\phi$ and activity
$z$ which fulfills a Ruelle bound. If for
a.a. $\ga$ the kernel
\begin{equation}\label{CoGibbsposdef}
e^{-\frac12 E(x,\ga)}e^{-\frac12 E(y,\ga)}z (1-e^{-\phi(x-y)})+(1-c) \delta(x-y)
\end{equation}
is positive definite then the coercivity inequality \eqref{poi} for $L$
with constant $c$ holds for all $F \in \FC$.
\end{corollary}

\subsection{Potentials increasing the spectral gap}\label{subsecinc}

For the Poisson point process, i.e. the Gibbs measure for the
potential $\phi=0$, one
has the spectral gap $c=1$, which follows also directly from condition~\eqref{Gibbsposdef}.
In order to prove condition~\eqref{CoGibbsposdef} for $c =1$ it is
obviously sufficient to prove
non-negativity (for a.a. $\ga$) of the expression for all $\psi \in \mathcal{C}_0^\infty(\R^d)$
\begin{equation}
 \int_{\R^d} \int_{\R^d} e^{-\frac12E(x,\ga)}e^{-\frac12E(y,\ga)}(1-e^{-\phi(x-y)})\psi(y)\psi(x) dxdy.
\end{equation}
and considering this a bilinear form in $e^{-\frac12E(x,\ga)}\psi(x)$
and recalling that due to Ruelle bound and regularity the latter
function is integrable,
one is lead to the following sufficient
condition
\begin{equation} \label{posmeyer}
 \int_{\R^d} (1-e^{-\phi(x)})\psi\ast\psi(x) dx,
\end{equation}
where $\psi\ast\psi$ denotes the convolution
of $\psi$ with $\psi$. Recalling the following definition
\begin{define}
A locally bounded measurable function $u: \R^d \mapsto \mathbb{C}$ is called positive definite if
for all $\psi \in \mathcal{C}_0^\infty(\R^d)$ holds
\[
\int_{\R^d}\int_{\R^d} u(x) \psi  \ast \psi(x) dx  \geq 0
\]
and
$u(0)\leq 1.$
\end{define}
As $1-e^{-\phi}$ is bounded, condition~\eqref{posmeyer} means that $f:
x \mapsto 1-e^{-\phi}$ is a positive definite function.

\begin{remark}
Note that the condition~\eqref{posmeyer} does not depend on $z$.
\end{remark}

To apply this
condition we now investigate if there exists any potential
$\phi$ such that $f$ is positive definite and $\phi$ fulfills the conditions
guaranteeing the existence of a Gibbs measure, namely (SS) and (R).

\begin{theorem}\label{SuffCondG}
Let $f$ be a continuous positive definite function which is (R). Define
\begin{equation}
\label{defpot}
\phi := - \ln(1-f).
\end{equation}
Then $\phi$ fulfills \eqref{posmeyer} and is (SS) and (R).
For every Gibbs measure $\mu$ for the potential $\phi$ and \textbf{ for
  any} activity $z$ which fulfills a Ruelle bound the associated generator $L$ of the
Glauber dynamics fulfills a coercivity inequality for $c  = 1$ and all
$F \in \FC$.
\end{theorem}

\noindent {\it Proof.}
Due to positive definiteness $|f(x)|\leq f(0) \leq 1$.
Defining for $x \in [-1,1]$ the function $h(x)=-\ln(1-x)$ one can write $\phi=h\circ f$.
First, we show that $\phi$ is regular. As$f$ is regular there exists an $\tilde{R}>0$ and a positive decreasing function $\varphi$
on $[0,+\infty)$ which fulfills
(\ref{eq4}) and such that $|f(x)|\le\varphi(|x|)$ for all  $x\in\R^d$ with $|x|\ge \tilde{R}$.
Note that for $x\in [-1,1/2]$ it holds that $|h(x)| \leq 2 x$. Choose
an $R\geq \tilde{R}$ such that $\varphi(R)\leq 1/2$.
Then for all $x \in \R^d$ with $|x|\geq R$ it holds $|f(x)|\leq 1/2$ and hence
\[
|\phi(x)| \leq 2f(x)\leq 2\varphi(|x|),
\]
which implies that $\phi$ is regular.

Second, we show that $\phi$ is superstable. One easily sees that
$h(x)\geq x+\One_{[2/3,1]}(x)(-\ln(1-x)-x)$. Shorthanding $g(x)=-\ln(1-x)-x$ one obtains
$\phi(x)\geq f(x)+\One_{[2/3,1]}(f(x))g(f(x)).$
Hence, $\phi$ is bigger then the sum of a positive definite function and a continuous function, which is
positive in 0, therefore due to Lemma~1.2 in \cite{Ru70} the potential $\phi$ is a superstable.
\quad $\square$\vspace{2mm}

We now try to understand the structure of potentials
fulfilling condition~\eqref{posmeyer}. For that let us  recall the following definition

\begin{define}
A generalized function (distribution) $u\in\D(\R^d)$ is called positive definite if for all
$\varphi\in\\mathcal{C}_0^{\infty}(\R^d)$
\begin{equation}
\la u,\tilde\varphi\star\varphi\rangle \geq 0
\end{equation}
holds, where $\tilde\varphi(x):= \overline{\varphi(-x)}$.
\end{define}

\begin{prop}
\label{propstruc}
Let $\phi$ be a potential fulfilling condition~\eqref{posmeyer} which is (S), (R),
and lower semi-continuous at zero.
Then it is of the from \eqref{defpot} and hence also (SS). Furthermore, $\phi$ is integrable, itself
positive definite in the sense of generalized functions, and
\begin{equation}
\label{growthphi}
\limsup_{x \downarrow 0} (\phi(x) +2 \ln(x)) < \infty
\end{equation}
\end{prop}

\noindent {\it Proof.} Let us define $f:=1-e^{-\phi}$ and show that
the function $f$ fulfils the conditions of Theorem~\ref{SuffCondG}.
As $\phi$ is stable it is non-negative in $0$ and hence $|f(0)| \leq 1$. Furthermore, $f$ is
lower semi-continuous at zero. Due to the positive
definiteness of $f$ one has that $f$ is continuous and
$|f(x)|\leq f(0) \leq 1$. One obtains the representation \eqref{defpot} by
inverting the definition of $f$. As in the proof of
Theorem~\ref{SuffCondG}
one can check that $f$ also fulfills (R).
Then Theorem~\ref{SuffCondG} implies that $\phi$ is
also (SS).

Using that $1-\cos(x) \geq \frac{x^2}{2}$ for small enough $x$,
$f$ is non-negative, the positive definiteness and $f(0)\leq 1$, we obtain that there
exists a constant $c>0$ such that $1-f(x) \geq c|x|^2$ for small
enough $x$. Hence $\phi(x)\leq -2\ln(|x|)-\ln(c)$.
As $\phi$ is bounded below and regular, it is integrable.

Writing again $\phi=h\circ f,$ we note that $h(x)=-\ln(1-x)=
\sum\limits_{n=1}^{\infty}\frac{x^n}{n}$ with radius of convergence $1$.
Approximate $\phi$ by the functions $\phi_{\delta}(x):=h\circ((1-\delta)f(x))$  for $0<\delta<1$.
Since $|(1-\delta)f(x)|< 1$ and $h$ has a Taylor series with non-negative
coefficients, for all $0<\delta<1$ the function $\phi_{\delta}$ is positive definite, cf. e.g.
\cite[Proposition~3.5.17]{Ja01}. As $h$ is monotone increasing $|\phi_{\delta}|\leq |\phi|$ and
the latter function is integrable. Hence $\phi_{\delta}$ is also positive definite in
the sense of generalized functions. Since $\phi_{\delta}$ converge pointwise to
$\phi$ for $\delta\to 0$ uniformly bounded by $\phi$, by Lebesgue's dominated convergence
$\phi$ is also positive definite in the sense of generalized functions.
\quad $\square$\vspace{2mm}

\subsection{Parameter dependence}

A typical question in statistical mechanics is to study the behavior of the system under change of a parameters. In the previous subsection we identify potentials which fulfill \eqref{posmeyer} for all $z$ and hence will show no phase transition even for large $z$. To investigate the temperature dependence we reintroduce the inverse temperature $\beta>0$ in to our consideration, that is we consider instead of $\phi$ the potential $\beta \phi$. We consider $\phi$ as fix and vary $\beta$ and $z$. The corresponding Papangelou intensity is $r(x,\gamma)=ze^{-\beta E(x)}$ and hence condition~\ref{posmeyer}
takes the form
\begin{equation} \label{posmeyerpa}
 \int_{\R^d} (1-e^{-\beta \phi(x)})\psi\ast\psi(x) dx.
\end{equation}
If \eqref{posmeyerpa} is positive for all $\psi \in \mathcal{C}_0^\infty(\R^d)$ then we say that
$\phi$ fulfills condition \eqref{posmeyerpa} for $\beta$.
Note, that the condition is independent of the activity $z$.

\begin{prop}
Let $\phi$ be a potential which fulfills condition~\eqref{posmeyer} for a $\bar{\beta}>0$ and is (S), (R),
and lower semi-continuous at zero.
Then $\phi$ fulfills condition~\eqref{posmeyer} for all $0 < \beta \leq \bar{\beta}$.
\end{prop}

\noindent {\it Proof.} Denote by $f:=1- e^{-\bar{\beta} \phi}$ the function considered in condition~\eqref{posmeyer}, which is positive
definite by assumption. One the one hand, it is easy to see that
$f_\beta(x):=1-e^{-\beta\phi(x)}$ are also continuous
and (R).  One the other hand,
 $f_{\beta}(x)=1-(1-f(x))^{\beta/\bar{\beta}}$ has a power series
 expansion $f_{\beta}(x)=\sum_{n=1}^{\infty}\frac{(-1)^{n+1}}{n!}\beta/\bar{\beta}(\beta/\bar{\beta}-1)\ldots
 (\beta/\bar{\beta}-n+1)(f(x))^n$ with radius of convergence $1$.
All the coefficients of the series are nonnegative, if $\beta/\bar{\beta}\le 1.$
Proceeding as in Proposition~\ref{propstruc}, one proves that $f_{\beta}$ is the pointwise
limit of positive definite functions. As $f_{\beta}$ is itself bounded and a limit of positive definite functions, it is a positive definite in the sense of functions.
\quad $\square$\vspace{2mm}

\subsection{Examples}

For concreteness we give a small collection of potentials which fulfills the
condition of Theorem~\ref{SuffCondG} to get a better feeling how such
potentials may look like. Especially interesting is that among them are
potentials, which have a non-trivial negative part.

\vspace{0.5cm}
\begin{tabular}{l|l|l}
\hspace{2cm}$\phi(x) $ &\hspace{1.5cm} $f(x)$ & Parameters\\[0.2cm] \hline & & \\
$-\ln(1-e^{-tx^2}\cos(ax)),$ & $e^{-tx^2}\cos(ax)$, & $t>0, a\in\R$\\[0.2cm]
$-\ln(1-e^{-t|x|}\cos(ax)),$ &  $e^{-t|x|}\cos(ax)$, &  $t>0, a\in\R$\\[0.2cm]
$\displaystyle -\ln\left(1-\frac{\cos(ax)}{1+\sigma^2x^2}\right),$  &
$\displaystyle \frac{1}{1+\sigma^2x^2}\cos(ax),$ & $\sigma>0,$ $a\in\R$\\[0.2cm]
$-\ln(1-(1-\frac{|x|}{a})1\!\!1_{[-a,a]}(x)\cos(bx)),$&
$(1-\frac{|x|}{a})1\!\!1_{[-a,a]}(x)\cos(bx),$ &   $a>0, b\in\R,$  \\
\end{tabular}

\vspace{0.2cm}

\noindent In all examples above one can exchange $\cos(ax)$ by $\displaystyle \frac{\sin(ax)}{ax}.$
 \vspace {0.5cm}

\noindent In the $d$-dimensional case we can give following examples:

\vspace{0.5cm}
\begin{tabular}{l|l|l}
\hspace{2cm}$\phi(x) $ &\hspace{1.5cm} $f(x)$ & Parameters\\[0.2cm] \hline & & \\
$-\ln(1-e^{-t|x|^2}\cos(a\cdot x))$ & $e^{-t|x|^2}\cos(a\cdot x)$ & $x\in\R^d, t>0, a\in\R^d$ \\[0.2cm]
$\displaystyle -\ln\left(1-e^{-t|x|^2}\prod\limits_{j=1}^d
\frac{\sin(a_jx_j)}{a_jx_j}\right)$ & $\displaystyle e^{-t|x|^2}\prod\limits_{j=1}^d
\frac{\sin(a_jx_j)}{a_jx_j}$ & $x\in\R^d, t>0$ \\[0.2cm]
 $-\ln\left( 1-(\frac{r}{|x|})^{n/2}J_{n/2}(r|x|)\right)$ &
 $(\frac{r}{|x|})^{n/2}J_{n/2}(r|x|)$& $r\geq 0$, $n > 2d-1$ \\[0.2cm]
$-\ln\left(1-\frac{2^{n/2}\Gamma(\frac{n+1}2)}{\sqrt\pi}\cdot\frac{t}{(|x|^2+t^2)^{\frac{n+1}2}}\right)$
&
$\frac{2^{n/2}\Gamma(\frac{n+1}2)}{\sqrt\pi}\cdot\frac{t}{(|x|^2+t^2)^{\frac{n+1}2}}$
& $t>0$, $n > d-1$\\[0.2cm]
\end{tabular}
where $J_{n/2}$ is the Bessel function of the first kind of order
$n/2$. One can multiply $f$ in any of the examples with factors of the
form $\cos(a\cdot x)$ and $\prod\limits_{j=1}^d
\frac{\sin(a_jx_j)}{a_jx_j}$.

All these examples are constructed by choosing a positive definite function $f$ and express
$\phi(x)=-\ln(1-f(x))$.

\subsection{High temperature and low densities}

In the previous subsection we considered potentials which
increase the spectral gap. Such potentials admit at most a logarithmic
singularity at zero. In this subsection we will show that
one may add a non-negative potential to these kind of potentials. However,
the constant $c$ in the coercivity inequality will decrease and the
spectral gap will depend on the activity of $z$.

\begin{theorem}
Let $\phi_1$ be bounded below, (R) and (S) and $\phi_2$ a potential fulfilling the
conditions of Theorem~\ref{SuffCondG}. Then for
every Gibbs measure $\mu$ for the potential $\phi_1+\phi_2$ and the
activity $z$, the associated generator $L$ of the
Glauber dynamics fulfills a coercivity inequality for the constant
\[
c = 1 - \left( \sup_{y \in \mathbb{R}^d}\right) \rho^{(1)}_\mu(y) \int_{\R^d} dx e^{-
  \phi_2(x)}|1-e^{-\phi_1(x)}|,
\]
where $\rho^{(1)}_\mu(y):= \int_{\Gamma} e^{-E(y|\gamma)}
\mu(d\gamma)$  is the first correlation function.
\end{theorem}

\noindent {\it Proof.} The main idea is to apply condition~\ref{CoGibbsposdef} directly.
In order to prove positive definiteness of the kernel \eqref{CoGibbsposdef} one has to prove non-negativity
of the following expression for all $\psi \in \mathcal{C}_0^\infty(\R^d)$
\begin{equation}
\label{posgen}
\int_{\R^d} dx \int_{\R^d} dy \psi(x) \psi(y) \left[
e^{-\frac12 E(x,\ga)}e^{-\frac12 E(y,\ga)}z (1-e^{-\phi(x-y)})+(1-c) \delta(x-y)\right]
\end{equation}
Rewriting
\[
1-e^{- \phi}= 1- e^{- \phi_2} + e^{- \phi_2} (1-e^{- \phi_1}).
\]
the first part of  \eqref{CoGibbsposdef} takes the form
\[
e^{-\frac12 E(x,\ga)}e^{-\frac12 E(y,\ga)}z (1-e^{-\phi_2(x-y)})+
e^{-\frac12 E(x,\ga)}e^{-\frac12 E(y,\ga)}z e^{- \phi_2(x-y)}(1-e^{-\phi_1(x-y)})
\]
As in the beginning of Subsection~\ref{subsecinc} the first summand is a
positive definite due to the assumptions on $\phi_2$. The second
summand can be bounded as follows
\begin{eqnarray*}
\lefteqn{\int_{\R^d} dx \int_{\R^d} dy \psi(x)e^{-\frac12 E(x,\ga)}\psi(y)
e^{-\frac12 E(y,\ga)}z e^{- \phi_2(x-y)}(1-e^{-\phi_1(x-y)})}\\
&\geq& - z \int_{\R^d} dx e^{-\phi_2(x)}|1-e^{-\phi_1(x)}|
\int_{\R^d} dy |\psi|(x+y)e^{-\frac12 E(x+y,\ga)}|\psi(y)| e^{-\frac12 E(y,\ga)}
\end{eqnarray*}
Applying Cauchy-Schwarz inequality to the last factor one obtains
\begin{eqnarray*}
\lefteqn{z \int_{\R^d} dy |\psi|(x+y)e^{-\frac12 E(x+y,\ga)}|\psi|(y) e^{-\frac12 E(y,\ga)}}\\
&\leq& \int_{\R^d} dy \rho^{(1)}(y) \psi^2(y).
\end{eqnarray*}
Summarizing \eqref{posgen} can be bounded below by
\begin{equation}
\int_{\R^d} dy \left[- \rho^{(1)}(y) \int_{\R^d} dx   e^{-\phi_2(x)}|1-e^{-\phi_1(x)}| + (1-c) \right]  \psi^2(y)
\end{equation}
which is non-negative if and only if the bracket is non-negative.
\quad $\square$\vspace{2mm}

\begin{center}
{\bf Acknowledgements}\end{center}

We thank Prof. Ludwig Streit for valuable discussion and his suggestion to
invert the positivity condition which made this work possible.
N.O gratefully acknowledge the financial support of the SFB~701
international graduate colleague~1132 ``Stochastics and real world models''.
T.K. thanks the project ``Ricostruzione di processi di  punto con  densit\`a
                        e funzioni di  correlazione assegnate'' 
                        for its generous support

\end{document}